\documentclass[preprint, preprintnumbers,amsmath,amssymb, nofootinbib]{revtex4}

\usepackage{graphicx}
\usepackage{dcolumn}
\usepackage{bm}
\usepackage{epsfig}
\usepackage{enumerate}

\usepackage{color}

\usepackage{amsmath}
\usepackage{amssymb}

\def\fun#1#2{\lower3.6pt\vbox{\baselineskip0pt\lineskip.9pt
  \ialign{$\mathsurround=0pt#1\hfil##\hfil$\crcr#2\crcr\sim\crcr}}}

\input epsf

\newcommand{\be}{\begin{equation}}
\newcommand{\ee}{\end{equation}}
\newcommand{\bi}{\begin{itemize}}
\newcommand{\ei}{\end{itemize}}
\newcommand{\bea}{\begin{eqnarray}}
\newcommand{\eea}{\end{eqnarray}}

\begin{document}
\global\long\def\met{\not{\!{\rm E}}_{T}}

\preprint{ANL-HEP-PR-11-64, MADPH-11-1576}
\vspace*{1cm}

\title{The dark side of the Higgs boson}

\vspace*{0.2cm}

\author{
\vspace{0.5cm} 
Ian Low$^{a,b}$, Pedro Schwaller$^{a,c}$, Gabe Shaughnessy$^{a,b,d}$ and Carlos E.~M.~Wagner$^{a,e,f}$ }
\affiliation{
\vspace*{.5cm}
$^a$ \mbox{High Energy Physics Division, Argonne National Laboratory, Argonne, IL 60439}\\
$^b$ \mbox{Department of Physics and Astronomy, Northwestern University, Evanston, IL 60208} \\
$^c$  \mbox{Department of Physics, University of Illinois, Chicago, IL 60607}\\
$^d$  \mbox{Department of Physics, University of Wisconsin, Madison, WI 53706}\\
$^e$  \mbox{Enrico Fermi Institute, University of Chicago, Chicago, IL 60637}\\
$^f$  \mbox{Kavli Institute for Cosmological Physics, University of Chicago, Chicago, IL 60637}
\vspace*{0.8cm}}

\begin{abstract}
\vspace*{0.5cm}
Current limits from the Large Hadron Collider exclude a standard model-like Higgs mass above 150 GeV, by placing an upper bound on the Higgs production rate. We emphasize that, alternatively, the limit could be interpreted as a lower bound on the total decay width of the Higgs boson. If the invisible decay width of the Higgs is of the same order as the visible decay width, a heavy Higgs boson could be consistent with null results from current searches.  We propose a method to infer the invisible decay of the Higgs by using the width of the measured $h\to ZZ\to 4 \ell$ lineshape, and study the effect on the width extraction due to a reduced signal strength.  Assuming the invisible decay product is the dark matter, we show that minimal models are tightly constrained by limits from Higgs searches at the LHC and direct detection experiments of dark matter, unless  the relic density constraint is relaxed.

\end{abstract}

\maketitle

\section{Introduction}\label{sec:into}

    High energy physics experiments have validated the Standard Model (SM) description at a high level of accuracy. 
These tests of the SM, however, have been restricted to the gauge sector of the theory. Very little is known about
the Higgs sector, related to the breakdown of the electroweak symmetry and the generation of mass of elementary
particles. Searches for the Higgs bosons at the Large Hadron Collider (LHC) are reaching maturity and it is expected that, independently of
the Higgs mass, if a SM-like Higgs  boson is present in the spectrum, the LHC will find evidence of it in the near
future. Indeed, data collected at the LHC so far are already excluding a SM Higgs mass above 150 GeV \cite{ATLAS-CONF-2011-112}.

The exclusion limit for a SM Higgs boson is presented in terms of an upper limit on the production cross section of the Higgs boson at the LHC; a particular value of the Higgs mass is considered excluded when, at the 95\% confidence level, the upper limit of the cross section reaches that expected of a SM Higgs boson. This way of presenting the exclusion limit is  well-motivated, since the limit is often derived from searches in many different channels, among which the production cross section is the universal strength modifier.

However, it is important to recall that what was actually measured in each search channel is the event rate, which is the product of the production cross section and the decay branching fraction. There are in fact two universal strength modifiers in the event rate across all search channels: the production cross section and the total width of the Higgs boson. Assuming  the production cross section is not affected, null results from the Higgs boson searches at the LHC could very well be interpreted as a {\em lower} limit on the total width of the Higgs.

New physics can affect interpretations of current Higgs exclusion limits in a significant way: it could modify the Higgs production cross section and/or the Higgs decay branching ratios.  An important example is the presence of new light colored particles which couple to the Higgs sector with a comparable strength to the top Yukawa coupling. Such new colored particles could alter the Higgs production cross section significantly in the gluon fusion channel, which is the dominant production channel at the LHC. This possibility is well-studied in the literature \cite{Djouadi:1998az, Low:2009nj}. To evade the exclusion limit, the Higgs production in the gluon fusion channel must be reduced from the SM expectation, pointing to scenarios where the Higgs mass is less fine-tuned \cite{Low:2009nj}. Another example of decreasing the production cross section is to induce mixing between the Higgs boson and a neutral scalar.\footnote{Such an effect will also change the width, but not the individual branching fractions as it suppresses all partial decay widths to the SM~\cite{singscalar,Barger:2006dh}.  Determining the pattern of the suppression for various modes can help determine the type of neutral scalar involved in the mixing~\cite{Barger:2009me}.}  The other possibility of reducing the branching ratios of Higgs decays in the relevant search channels  can be achieved by, for example, increasing the decay to competing SM modes such as $b\bar b$ in the light mass region \cite{Carena:2011fc}.  It is also possible that the Higgs has a larger than expected total width, which would then reduce the branching ratios universally in all decay channels. Such a scenario arises naturally when the Higgs boson couples to quasi-stable neutral particles with a mass that is smaller than half of the Higgs mass, in which case the Higgs invisible decay width may be of the same order or even larger than the visible decay width, thereby reducing the branching ratios into visible matter~\cite{singscalar,Hinvisible}.

In other words, the current Higgs search limit could be a hint on the ``dark side'' of the Higgs boson, suggesting a large invisible decay width.  Searches for Higgs particles decaying invisibly were performed at LEP and have been investigated at the LHC in both the associated production with vector bosons as well as in the vector boson fusion (VBF) channels. In the VBF case, it was suggested that a 14 TeV LHC is capable of probing the existence of such an invisible decay width with a modest integrated luminosity, of about 10~fb$^{-1}$ \cite{Eboli:2000ze}.  We will study an alternative method to infer the Higgs invisible width, by measuring the total width from the lineshape of Higgs decays into four leptons via two $Z$ bosons.   Since the experimental resolutions in the total invariant mass is at around 1 - 2 GeV \cite{Aad:2009wy,Ball:2007zza}, such a method becomes effective at large Higgs masses, above 190 GeV.  If the invisible particle the Higgs decays into is the dark matter, the amount of reduction required to satisfy the current search limit, as well as constraints from direct detection experiments, turns out to have interesting implications on the relic density of the dark matter.

This work is organized as follows.  In Section~\ref{sect:hinv}, we review the impacts on Higgs searches from the dark side of the Higgs, while in Section~\ref{sect:lineshape} we study the $4\ell$ lineshape measurement and extract the Higgs boson width.  We discuss the possible dark matter connections in Section~\ref{sect:DM} and provide concluding thoughts in Section~\ref{sect:conclude}.

\section{The dark side of the Higgs}\label{sect:hinv}
If there exists a light quasi-stable neutral particle that couples to the Higgs boson, the resulting decay may be seen as missing energy in a detector.  The presence of such a decay mode dilutes the strength of production and decay to visible states typically used for Higgs boson discovery. 

\begin{figure}[h]
\begin{center}
\includegraphics[scale=0.55, angle=0]{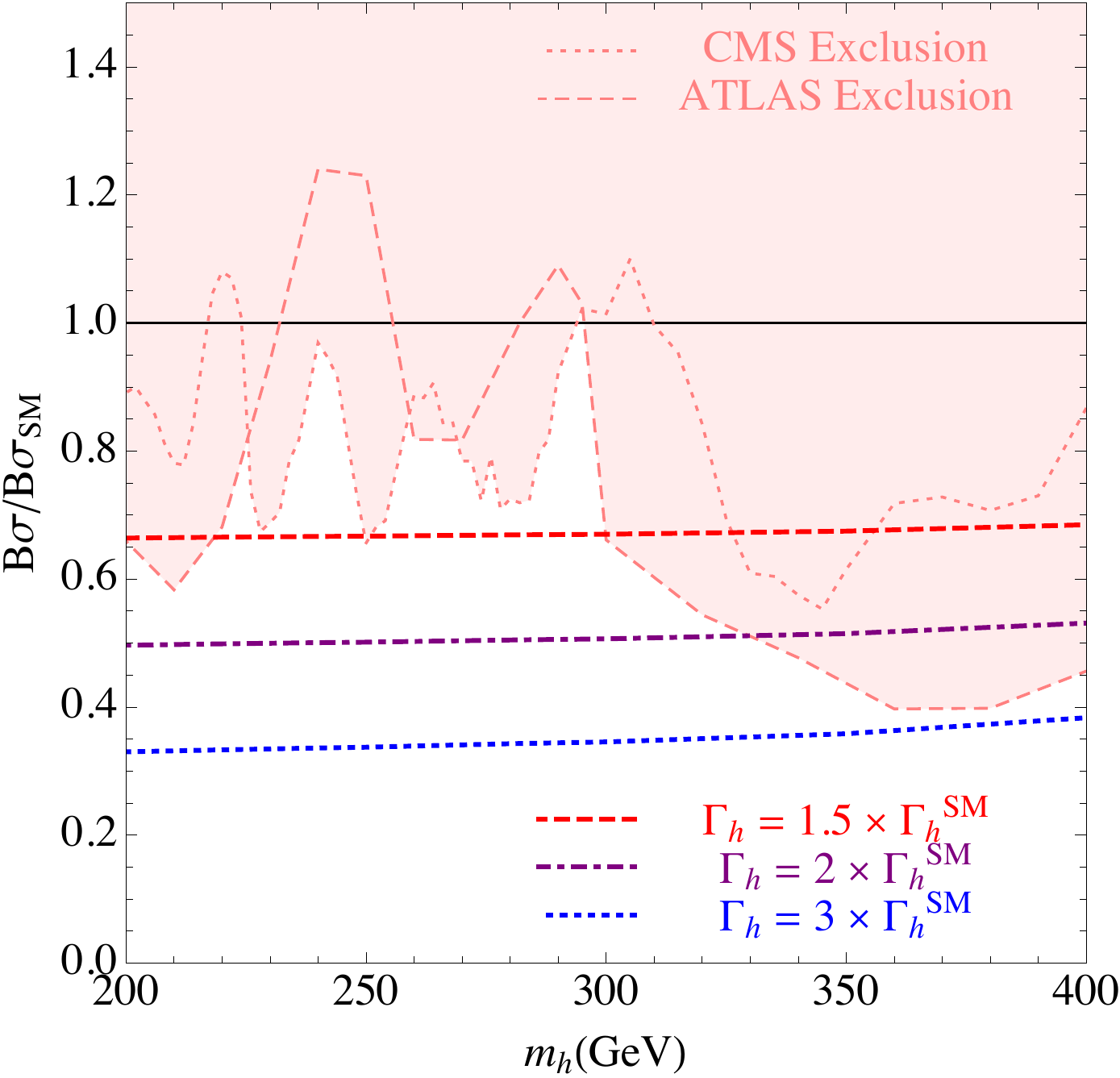}  
\caption{\it Current Higgs search limits at the LHC \cite{ATLAS-CONF-2011-112}, presented in terms of upper limits on the production cross section. We also show the equivalent suppression in the cross section from an increase in the total decay width of the Higgs that is 1.5, 2, and 3 times the SM total width.}
\label{fig:lhclimit}
\end{center}
\end{figure}

Assuming that the production strength of the Higgs boson is unchanged and the narrow-width approximation (NWA) is valid, the event rate of the Higgs signature in a particular channel $X_{\rm SM}$ is reduced by the fraction
\be
B\sigma(pp\to h\to X_{\rm SM}) = {\Gamma_{h_{\text SM}}\over \Gamma_{h_{\text SM}} + \Gamma(h\to X_{\text inv})}\times B\sigma^{\rm (SM)}(pp\to h\to  X_{\rm SM})\,,
\label{eq:sigred}
\ee
where $\Gamma_{h_{\rm SM}}$ is the SM Higgs boson total width, and $X_{\rm inv}$ are the non-SM invisible states the Higgs boson decays to.\footnote{Note that in the SM, the Higgs has a maximal invisible decay rate of around 1\% when $h\to ZZ\to \nu\bar\nu\nu\bar\nu$.} When the Higgs mass is heavy and the width becomes substantial, the NWA may not be valid. Therefore, in our study we calculate the suppression in the event rate using the full Breit-Wigner propagator.  Deviations from the NWA are found to be ${\cal O}(15\%)$ for $m_h=500$ GeV, but quite small for masses in the 200 GeV range.  In order to be consistent with the null result from current Higgs searches, the reduction factor in the event rate is generally required to be ${\cal O}$(50\%) or larger \cite{ATLAS-CONF-2011-112}, which suggests a dark side of the Higgs that is comparable to the visible side:
\be
\Gamma(h\to X_{\text inv}) \ \agt \ \Gamma(h\to X_{\text SM}) \ .
\ee
In Fig.~\ref{fig:lhclimit} we show the suppression necessary to be consistent with LHC search results from ATLAS and CMS collaborations.

The invisible width of the Higgs could be probed directly at the LHC by searching for $W/Z$ plus large missing energy in the associated production channel of the Higgs~\cite{Davoudiasl:2004aj}, or using the VBF channel by looking for two forward jets plus large missing energy~\cite{Davoudiasl:2004aj,Eboli:2000ze,Aad:2009wy}. However, at the LHC production rates in both channels are an order of magnitude smaller than that in the dominant gluon fusion channel. It is therefore desirable to look for additional ways to probe the dark side of the Higgs.

The total width of the Higgs can be measured from the lineshape in Higgs decays in the golden channel: $h\to ZZ\to 4\ell$. Because of the experimental resolution in total invariant mass, such a measurement can be made for Higgs masses above 190 GeV \cite{Aad:2009wy,Ball:2007zza}. In this mass range, the dominant decay modes of the SM Higgs boson are into $W$ and $Z$~boson pairs.
If one assumes a one-Higgs-doublet model where the electroweak symmetry is broken by the Higgs vacuum expectation value, as we will in this work, the neutral Higgs couplings to pairs of $WW$ and $ZZ$ are completely determined by the $SU(2)_L\times U(1)_Y$ gauge symmetry,
\be 
\label{eq:wzh}
\frac12 g^2 v\, h W_\mu^-W^{+\, \mu} + \frac14 \frac{g^2}{c_w^2} v\, h Z_\mu Z^\mu \ ,
\ee
where $v\approx 246$ GeV, $g$ is the $SU(2)_L$ coupling strength, and $c_w$ the cosine of the Weinberg mixing angle. 
Since in the heavy mass region the Higgs decay is dominated by $WW$ and $ZZ$ channels, the visible decay width of the Higgs is therefore well known below the top threshold. Above the top threshold, we will simply assume the Higgs coupling to the top quark is the same as in the SM. However, because the $t\bar t$ branching fraction is generically less than 20 \%, this is a rather weak assumption once experimental uncertainties in the total width measurement is taken into account.  In the end, the non-SM invisible width can be inferred from a measurement on the total width:
\be
\Gamma(h\to X_{\text inv}) = \Gamma_{\rm total} -  \Gamma(h\to WW+ZZ+t\bar{t}) \ .
\ee
These assumptions may be cross checked by measuring the Higgs couplings to $WW, ZZ$, and $t\bar{t}$ in the heavy mass region. Earlier studies on extracting Higgs couplings mostly concentrate on the low mass region \cite{Duhrssen:2004cv}. A recent study \cite{Logan:2011ey} did consider extracting Higgs couplings at 190 GeV by including using a similar measurement on the total width in the $4\ell$ lineshape.

Note that a measurement of the total Higgs boson width below the SM expectation would indicate either a suppression in the coupling (e.g.  by mixing), or a singlet acting as a ``Higgs imposter''.  In either case, the relative decay rates can shed light on whether the singlet mixes with the real Higgs boson~\cite{Barger:2009me} or is an imposter~\cite{Low:2010jp}.

\section{$4\ell$ lineshape}\label{sect:lineshape}

The $h\to ZZ\to 4\ell$ channel is often considered the ``gold-plated'' mode for discovering the Higgs boson at the LHC.  Due to the low backgrounds dominated by continuum $ZZ$ production, and a well measured final-state, the signal is quite clean and can be easily isolated from the background.  The $4\ell$ final state offers an opportunity to measure the Higgs boson mass very well, with uncertainty in the 1-0.1\% range.  Moreover, the width can be extracted from the $M_{4\ell}$ lineshape.  The  experimental resolution of the lepton momenta broadens the $4\ell$ lineshape, which makes this measurement impossible for light Higgs bosons, where the Higgs width is a few MeV.  However, for masses above 190 GeV, the lineshape is sensitive to the Higgs width within the SM.

The Higgs boson width extraction from the $4\ell$ lineshape has been studied in several CMS analyses \cite{CMSNote2006_107,CMSNote2006_136}. However, these studies all assumed a SM width, while our motivation calls for a larger total width with a reduced overall normalization in the lineshape. In this section we study the effect of the diluted event rate in the $4\ell$ channel due to a large invisible width on the sensitivity of extraction of the total width. More explicitly, we consider a range of Higgs masses between 200 and 500 GeV with widths between $1-3$ times the SM Higgs boson width, calculated using the HDECAY package~\cite{HDECAY}.

To analyze the LHC sensitivity beyond the aforementioned CMS studies, we generate 50000 events in $h\to ZZ\to 4\ell$ channel in Madgraph~\cite{Maltoni:2002qb} and assume the background subtraction can be done cleanly as this is a precision measurement, only made after first discovering the Higgs boson.  That said, we do include the increase in uncertainty after background subtraction.  Using the code ALPGEN~\cite{Mangano:2002ea}, we compute the irreducible background from continuum $ZZ$ production and reducible backgrounds from $t\bar t$ production and $Z b\bar b$ production, where the heavy flavor decays produce an isolated muon.  In practice, after cuts, we find the continuum $ZZ$ production dominates the background sample and peaks near $M_{ZZ}=200$ GeV.  We therefore concentrate on the irreducible background and apply a K-factor of $K_{ZZ}=1.6$ as calculated in MCFM~\cite{Campbell:1999ah}.  

The experimental broadening of the lineshape is estimated by generating events with a vanishing Higgs width and smearing the final state lepton momenta according to the experimental resolution in the electron and muon channel. To be precise, we use~\cite{Ball:2007zza}
\begin{align}
\left({\Delta p \over p}\right)_{\mu} &= 0.84\% \oplus 1\% \left({p_T\over 100 ~{\rm GeV}}\right), \\
\left({\Delta p\over p}\right)_{e} &= {2.8\% \over \sqrt {p/\text{GeV}}}\oplus 12.4\% {{\rm GeV}\over p} \oplus 0.26\%\,,
\end{align}
where $\oplus$ indicates that the errors are added in quadrature. The broadening of the lineshape is then obtained by fitting the result to a Gaussian distribution. 

The shape of the measured invariant mass distribution, $M_{4\ell}\equiv \sqrt{\hat s}$ is  described by a convolution
\begin{align}
	\frac{d\sigma}{d M_{4\ell}} & =\int dM^\prime \frac{d\sigma_{\rm BW}(\sqrt{\hat s}-M^\prime) }{dM_{4\ell}} \frac{d\sigma_{\rm Gauss}(M^\prime)}{dM_{4\ell}}\,
\end{align}
where the first term is the physical Breit Wigner shape of the Higgs resonance:
\begin{equation}
\frac{d\sigma_{\rm BW}(\sqrt{\hat s}) }{dM_{4\ell}} = { {\hat s}^{3/2} \sqrt{1-4 x_Z}(1-4 x_Z+12 x_Z^2)\over  ((\hat s-M_h^2)^2+M_h^2 \Gamma_h^2)}\,,
\end{equation}
and is found after fitting the $M_{\ell\ell}$ distribution with no smearing; here $x_Z \equiv M_Z^2/\hat{s}$.   The experimental broadening term is the gaussian distribution:
\begin{equation}
\frac{d\sigma_{\rm Gauss}(M^\prime) }{dM_{4\ell}} ={1\over \sqrt{2 \pi} \sigma_{\rm exp}}  e^{-{{M^\prime}^2 \over 2\sigma_{\rm exp}^2}}\,,
\end{equation}
where $\sigma_{\rm exp}$ is the channel dependent experimental broadening on the lineshape.  Using this procedure, it is possible to accurately measure the width of the Higgs boson down to masses of 200~GeV, where it is of the same order as the experimental resolution. 
This lineshape is modified by radiative corrections~\cite{Papavassiliou:1997pb} and by final state radiation. Both corrections do not affect our analysis in a significant way, and are therefore neglected in our analysis. 

In our analysis we utilize both the $4\mu$ and  $e^+e^-\mu^+\mu^-$ decay of the Higgs, which we discuss in turn.  The $4\mu$ channel benefits from a very clean muon identification and the absence of  $j,\gamma\to e$ fakes. The $e^+e^-\mu^+\mu^-$ channel benefits from having twice the signal rate compared with the $4\mu$ channel and a slightly better momentum resolution for hard electrons.

In the $4\mu$ channel, we follow the cuts outlined in Ref.~\cite{CMSNote2006_107} by requiring exactly four tagged muons with $|\eta| < 2.4$ and 
\bea
p_T(\mu_1)&>& 15\text{ GeV}, \quad p_T(\mu_2)> 15\text{ GeV}\\
p_T(\mu_3) &>& 12 \text{ GeV}, \quad p_T(\mu_4) > 9 \text{ GeV}
\eea
where $\mu_i$ are $p$-ordered muons.  Additionally, we require that at least one pair of opposite sign muons reconstruct the $Z$-boson,
\be
70\text{ GeV} < M_{\mu^+\mu^-} < 100\text{ GeV}.
\ee
The $4\mu$ lineshape does have a combinatorial background for the individual reconstruction of the $Z$-bosons, but selection efficiency is good.  

\begin{figure}[t]
\begin{center}
\includegraphics[scale=0.45, angle=0]{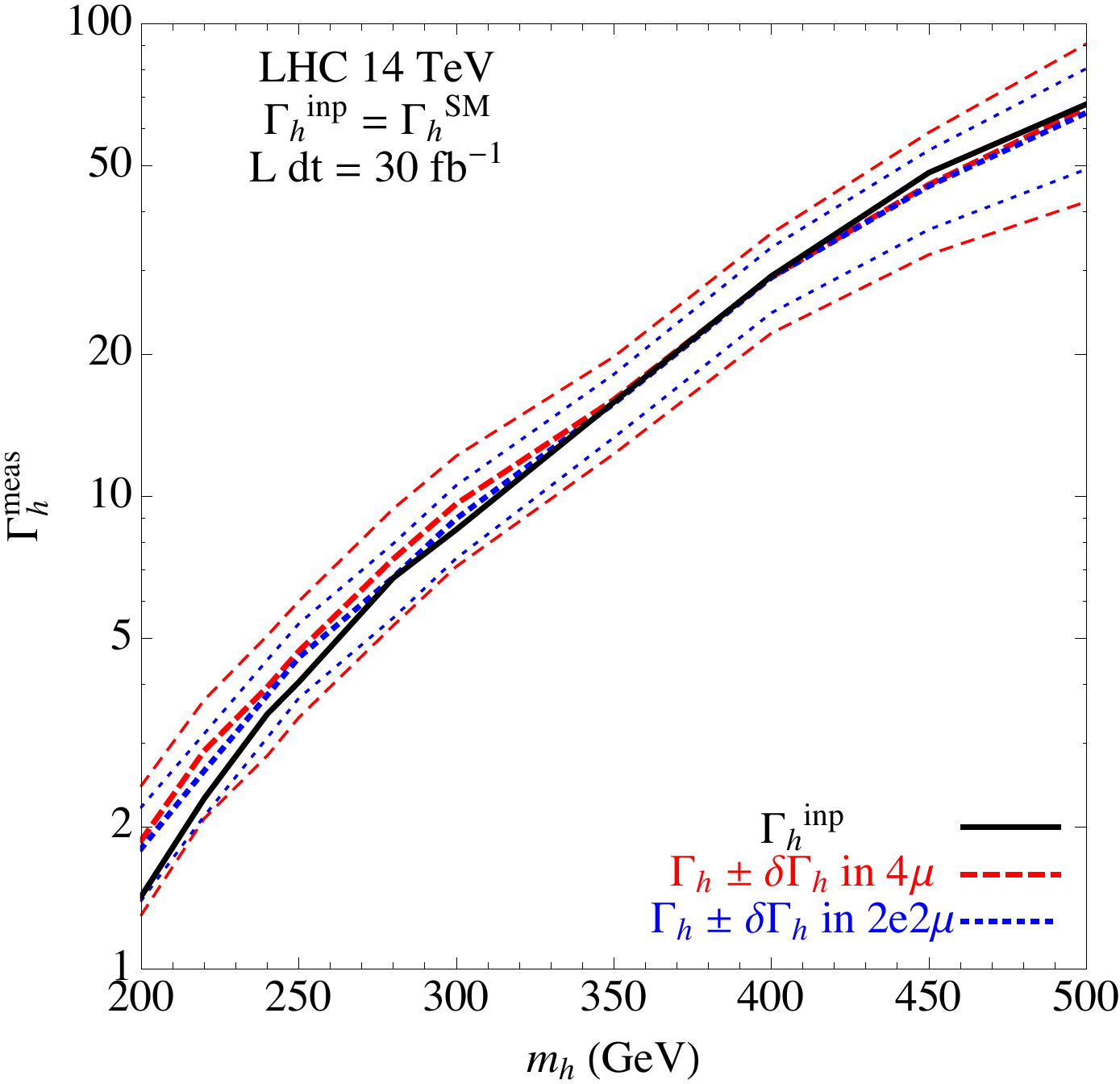}  
\includegraphics[scale=0.45, angle=0]{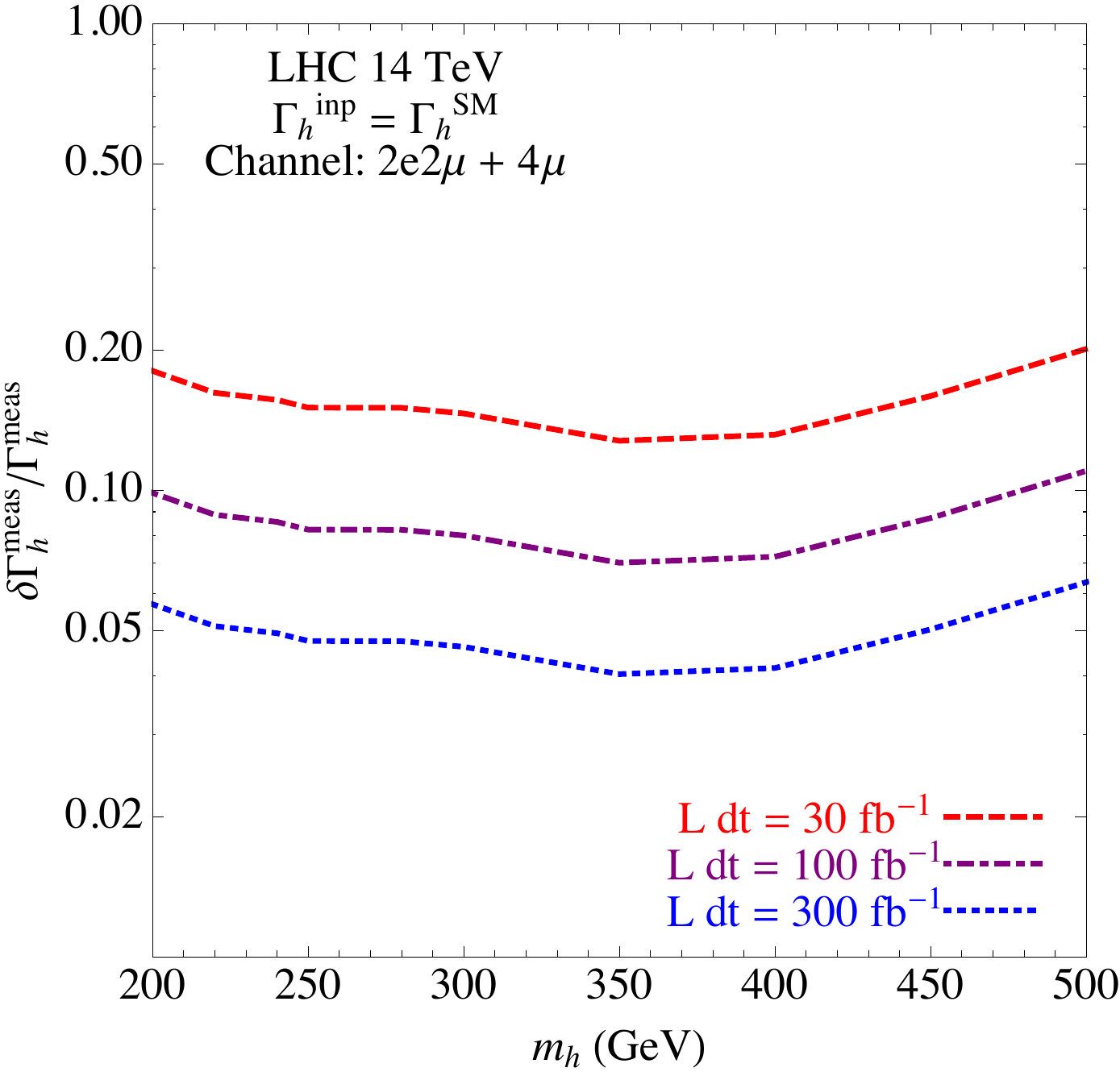}\\
\includegraphics[scale=0.45, angle=0]{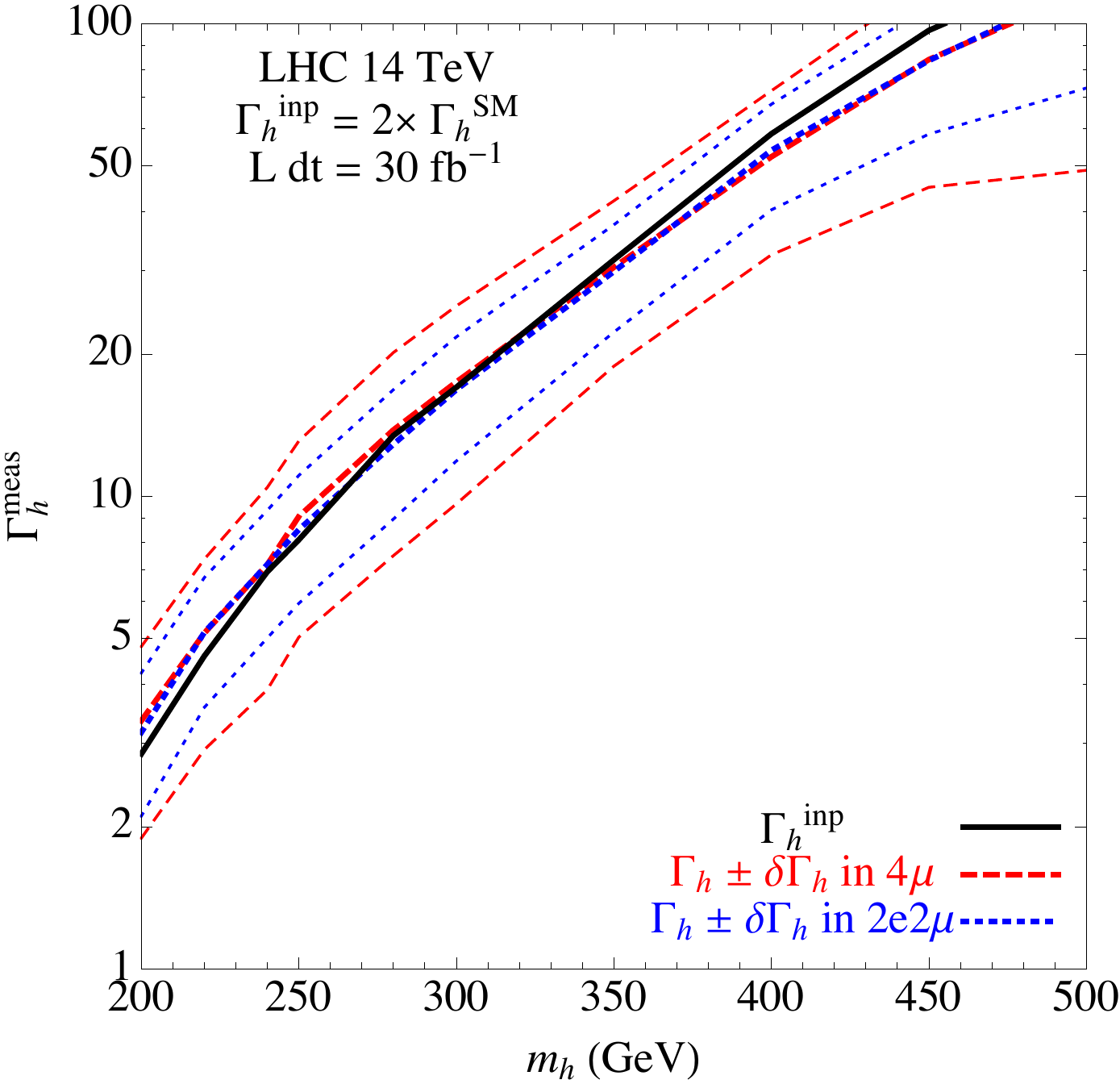}  
\includegraphics[scale=0.45, angle=0]{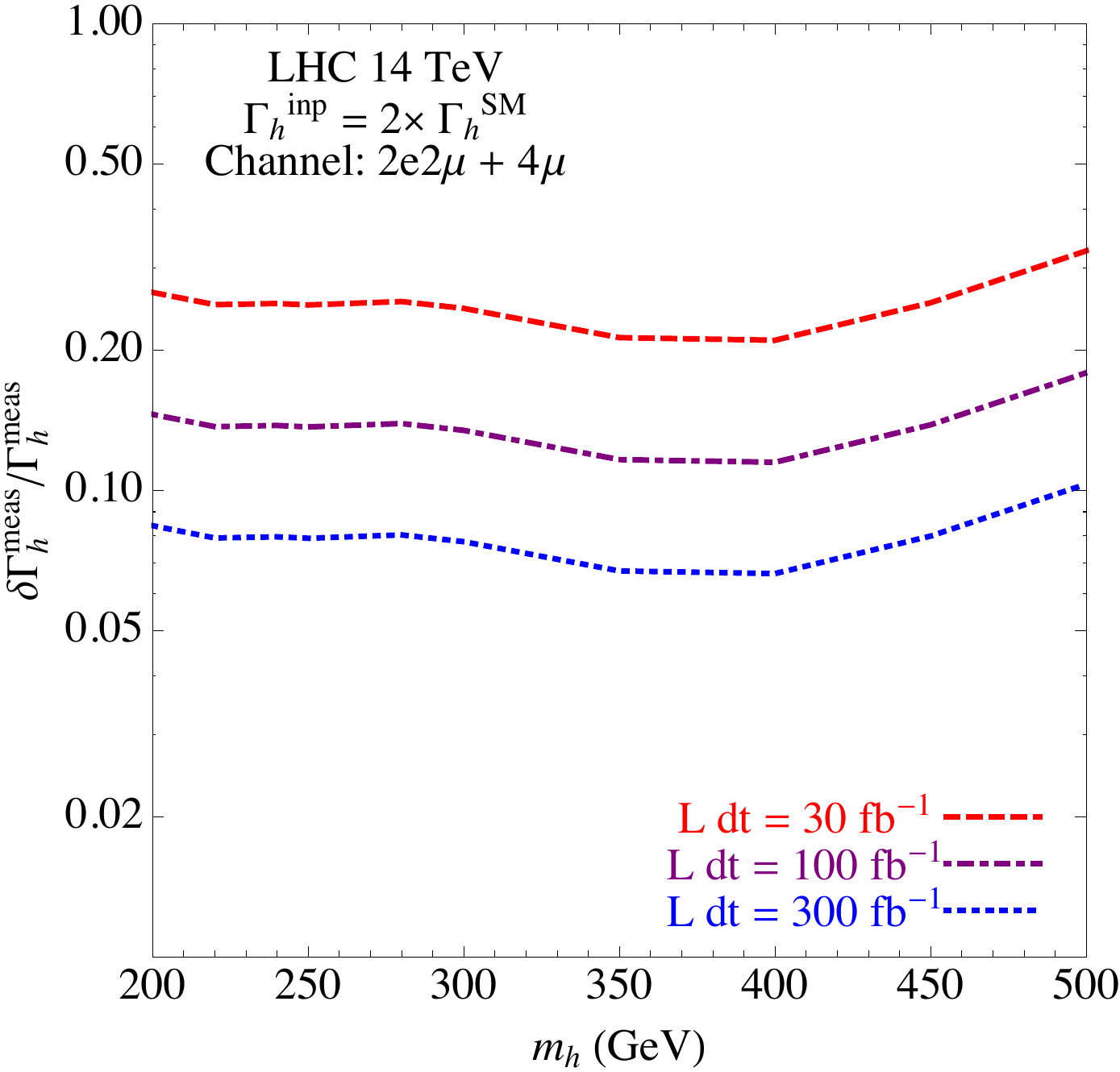}\\
\caption{\it The left panel shows measured Higgs width in range of Higgs masses in the $4\mu$ and $2e2\mu$ channels for 30 fb$^{-1}$ of integrated luminosity with $\Gamma_h^{\rm inp} = \Gamma_h^{\rm SM}$ and $2 \times \Gamma_h^{\rm SM}$.  The $1\sigma$ range is denoted by the light outer curves.  The input width, $\Gamma_h^{\rm inp}$, is denoted by the solid black curve. The right panel, on the other hand, shows $1\sigma$ fractional uncertainty extracted from the fit.  }
\label{fig:comb0}
\end{center}
\end{figure}

The cuts in the $2e2\mu$ channel are motivated by Ref.~\cite{CMSNote2006_136}  for a Higgs boson mass of 250 GeV, 
\bea
n^{\rm tagged}_{e^\pm}&=&2,\quad n^{\rm tagged}_{\mu^\pm}=2,\\
p_T(\ell_1) &>& 50\text{ GeV}, \quad p_T(\ell_2) > 35\text{ GeV},\\
p_T(\ell_3) &>& 25\text{ GeV}, \quad p_T(\ell_4) > 10\text{ GeV},\\
55\text{ GeV}&<&M_{e^+e^-,\mu^+\mu^-}<107\text{ GeV},
\eea
We find that with these cuts, we can efficiently reject the $Z+Z^*/\gamma^*$ as well as the reducible backgrounds $Zb\bar b$ and $t\bar t$ to a negligible level.

As emphasized already, previous studies on total width measurement in the $4\ell$ channel did not include the dilution effect from an enlarged invisible decay width. To demonstrate the impact of a reduced signal strength on the width measurement, in Fig.~\ref{fig:comb0} we show the extracted widths as well as the fractional uncertainties for an input width that is 1 times and 2 times the SM 
expectation. The measured width follows closely the input width and is largely  within the $1\sigma$ boundaries for both channels we consider and various luminosities.  By increasing the total width by a factor of two, the measurement uncertainty increases correspondingly as shown in the right panels.  Departure of the fit from the SM width beyond the uncertainty would suggest a breakdown of the assumption that the lineshape may be modeled as a convolution of the true lineshape and a gaussian kernel.  It is clear from these bands the $2e2\mu$ channel provides the most precise measurement of $\Gamma_h$, afforded by the better momentum resolution of the electrons. The fractional uncertainty of the measured Higgs total width at the $\sqrt{s}=14$ TeV LHC after combining the  $2e2\mu$ and $4\mu$ channels can be below 20\% for 30 fb$^{-1}$ of integrated luminosity over a broad mass range.

\begin{figure}[t]
\begin{center}
\includegraphics[scale=0.45, angle=0]{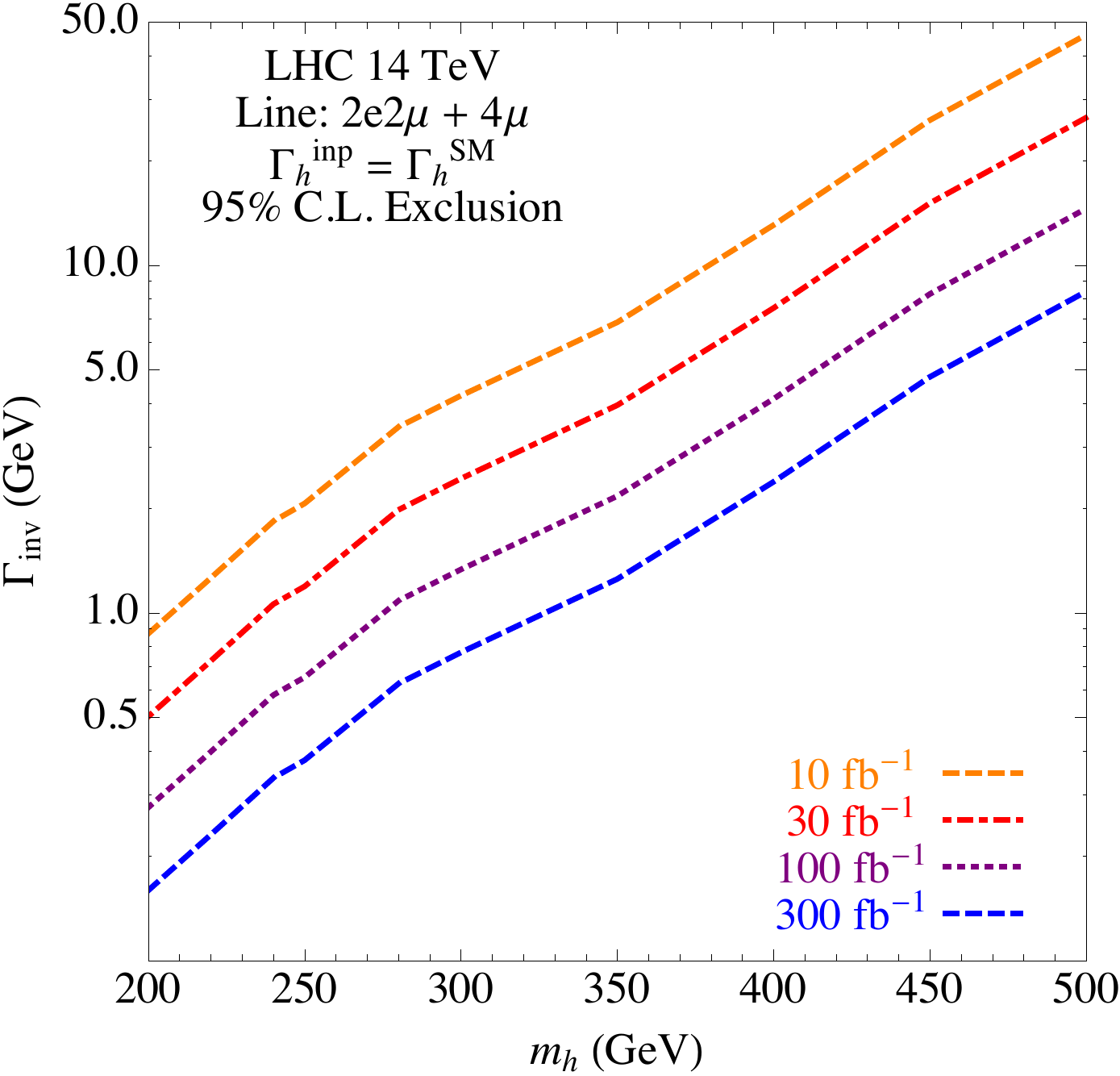}  
\includegraphics[scale=0.445, angle=0]{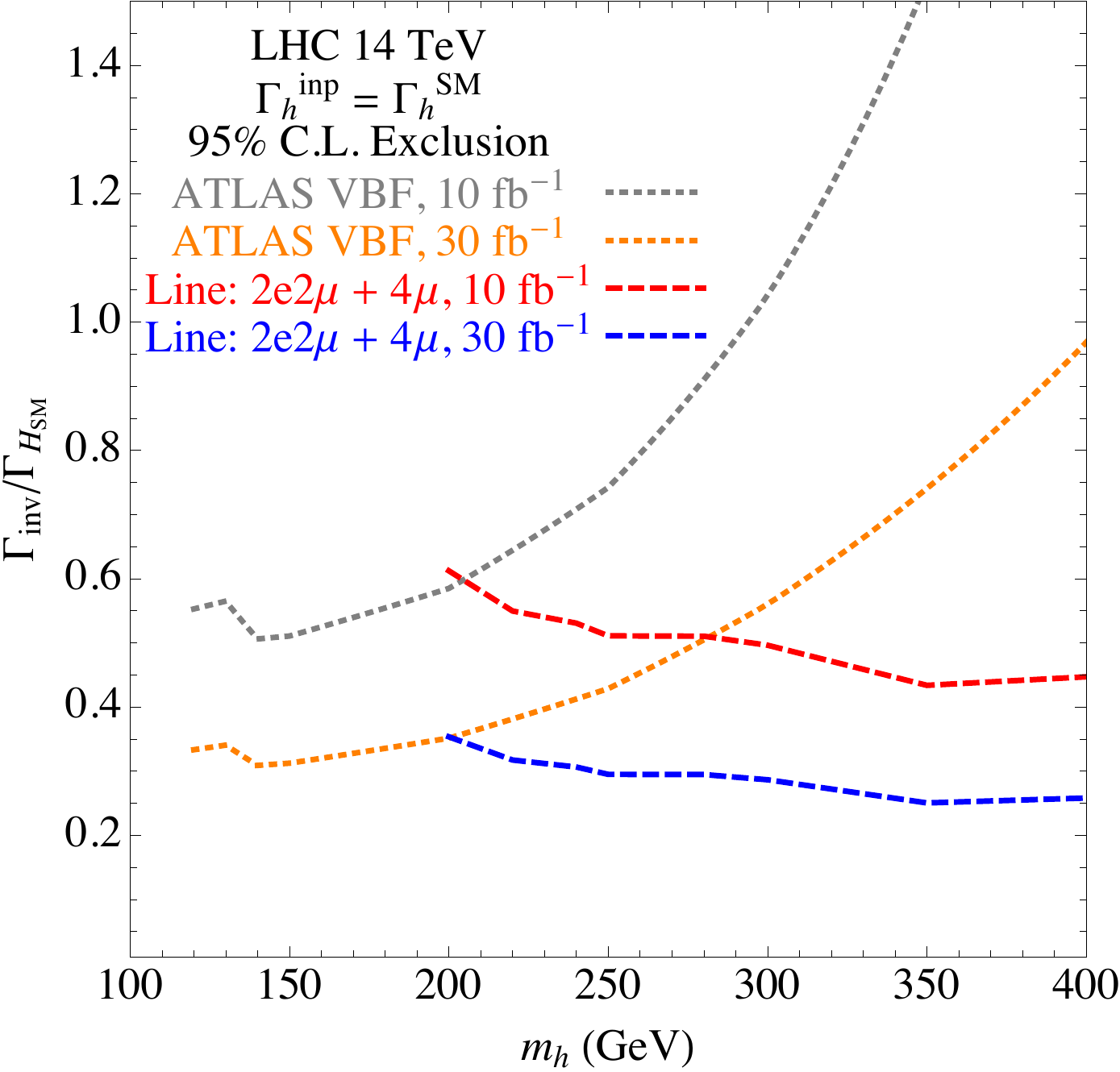}  
\caption{\it Assuming a SM input width, the uncertainty in width extraction can be converted to a 95\% C.L. reach in the invisible width, as shown in the left panel. The right panel shows comparison in invisible width measurements between direct probe via VBF  \cite{ATLHinv} and indirect probe via total width.  The direct and indirect probes are  complimentary to each other.}
\label{fig:reach}
\end{center}
\end{figure}

It is worth comparing the LHC reach of the invisible width through the VBF channel to the indirect measurement by the $4\ell$ lineshape, which is shown in Fig.~\ref{fig:reach}.  We see that the searches in VBF for missing energy and the lineshape measurement are very much complementary, with the VBF measurement being more sensitive for low Higgs masses, and the lineshape being more constraining for $m_h > 220$~GeV. However one should keep in mind that the VBF missing energy search directly probes invisible Higgs decays, while the lineshape measurement is only an indirect probe relying on the assumptions mentioned in Sect.~\ref{sect:hinv}.

\begin{figure}[h!]
\begin{center}
\includegraphics[scale=0.45, angle=0]{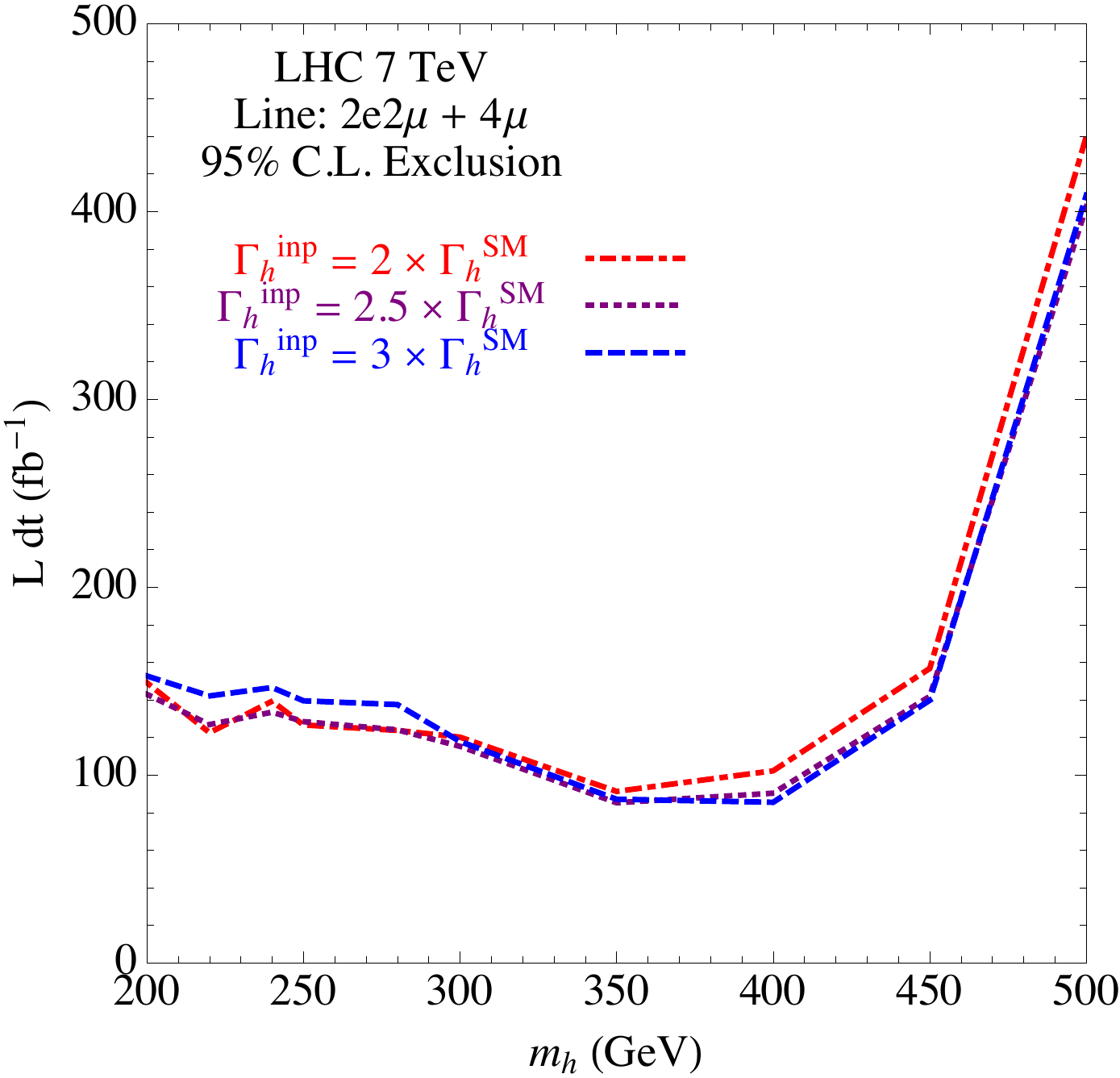}
\includegraphics[scale=0.45, angle=0]{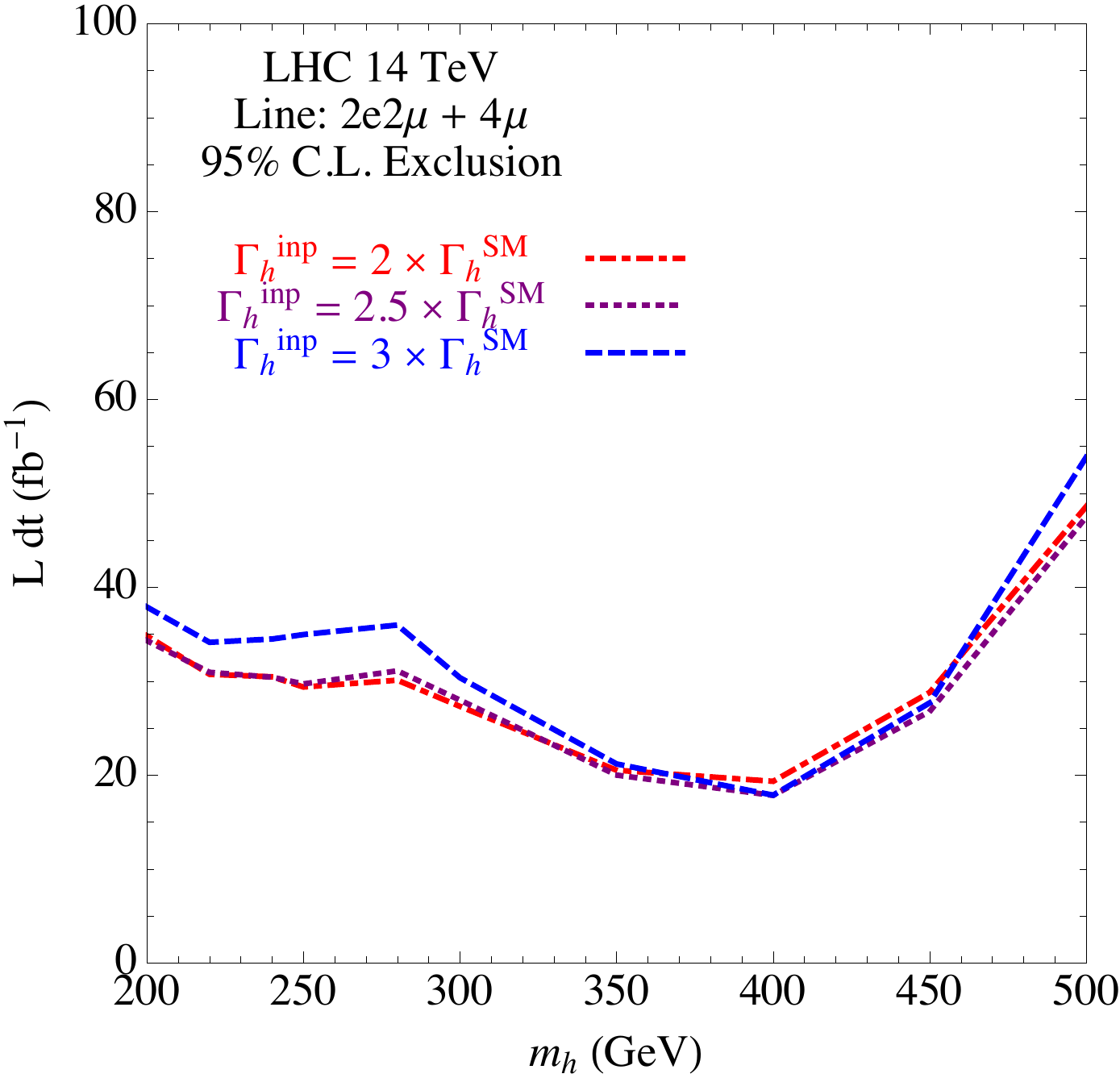}
\caption{\it Exclusion reach of a SM total decay width.  Comparison are made between total widths that are 2 times, 2.5 times, and 3 times the SM expectation.  }
\label{fig:mix1}
\end{center}
\end{figure}

In Fig.~\ref{fig:mix1} we show the luminosities needed for exclusion at the 95\% C.~L. of a SM total width in the $4\ell$ lineshape measurements, which would be a strong hint on the existence of a sizable invisible width. We consider scenarios when the input width is 2 times,  2.5 times, and 3 times the SM expectation at both the 7 TeV and 14 TeV LHC. More specifically, we simulate the luminosities at which
\be
\frac{\Gamma_{h}^{\rm meas} - \Gamma_{ h}^{\rm SM}}{\delta \Gamma_{h}^{\rm meas}} = 1.96 \ ,
\ee
where $\Gamma_{h}^{\rm meas}$ is the central value of the measured total width  and $\delta \Gamma_{h}^{\rm meas}$ is the $1\sigma$ uncertainty in the extraction.  As can be seen, at 7 TeV it would be extremely difficult to rule out a SM total width while at 14 TeV only a small amount of luminosity, less than 40 fb$^{-1}$, is needed unless the Higgs is heavier than 450 GeV. Somewhat surprisingly, having an input width larger than the SM expectation has little impact on the exclusion reach of a SM total width, at least in the three possibilities we considered. Although the uncertainty in the width extraction increases with a larger total width, as can be seen in Fig.~\ref{fig:comb0}, the difference between $\Gamma_{h}^{\rm meas}$ and $\Gamma_{ h}^{\rm SM}$ also increases in such a way that results in the behavior in Fig.~\ref{fig:mix1}. It would be interesting to see if this is still the case in more realistic simulations.

\begin{figure}[t]
\begin{center}
\includegraphics[scale=0.5, angle=0]{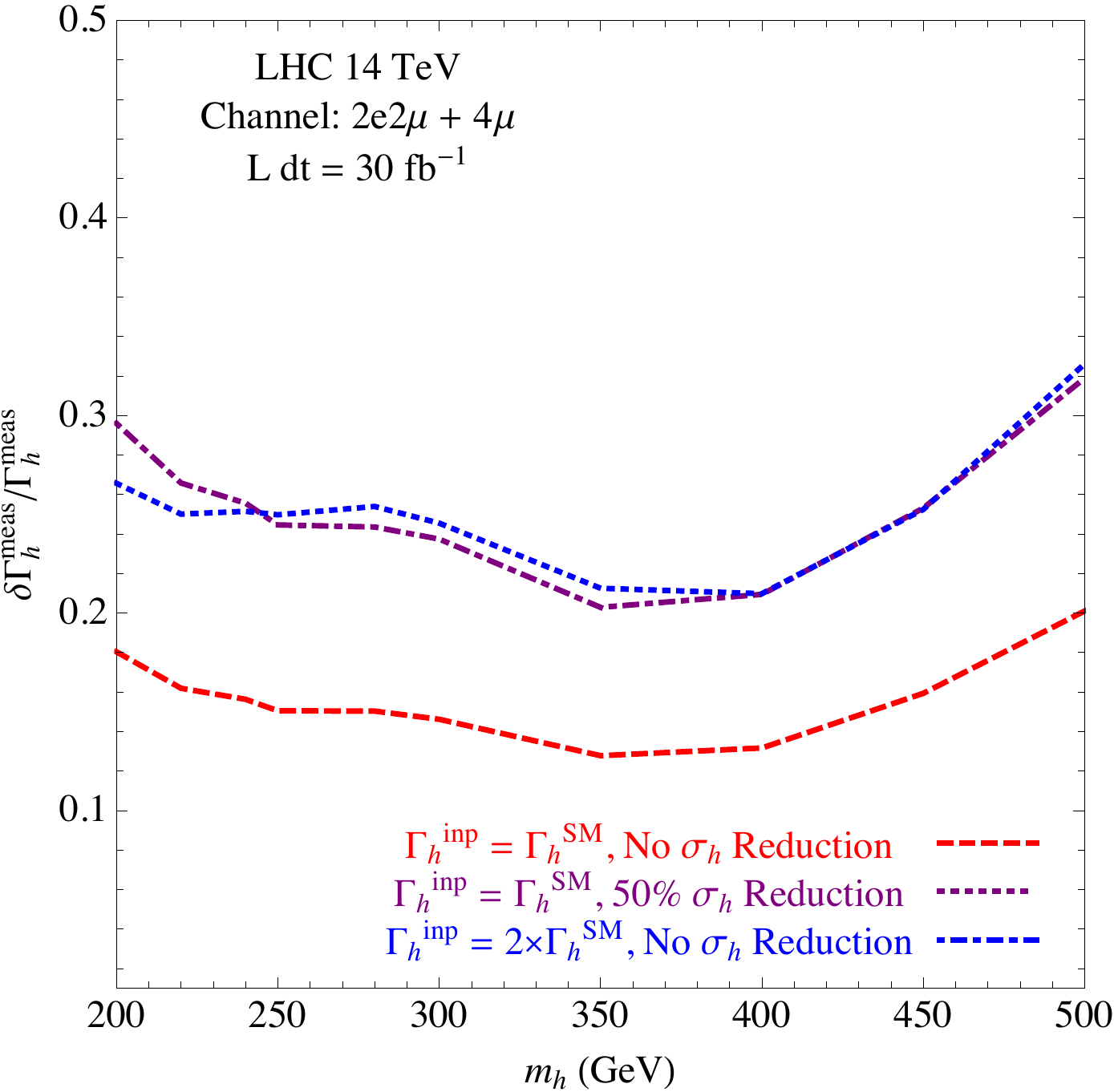}
\caption{\it Comparison of two different reduction mechanisms in the event rate.}
\label{fig:mix}
\end{center}
\end{figure}

Finally, it is worth recalling that a reduction in the event rate in a particular Higgs search channel could be due to i) a decrease in the production cross section or ii) an increase in the total width. In Fig.~\ref{fig:mix} we compare the fractional uncertainties in the width measurements from these two different causes when the event rate is only 50\% of that expected from the SM. In both cases the uncertainties are worse than that from a SM event rate, although there seems to be little difference between the two reduction mechanisms. However, since the total width in ii) is twice as large as that in i), the absolute uncertainty $\delta \Gamma_{h}^{\rm meas}$ in ii) is also twice as large as in i) due to the broadening of the lineshape. A measurement on the total width, when combined with a counting experiment measuring the event rate, could help disentangle the two reduction mechanisms. For example, a decrease in the event rate and a total width consistent with the SM would imply the reduction is due to a smaller than expected production cross section.

\section{Dark Matter Connections}\label{sect:DM}

If the Higgs boson decays to invisible particles, the presence of new quasi-stable states naturally leads to the question whether those particles could also be dark matter candidates. Assuming this is indeed the case, we study in this section implications from Higgs search limits, the observed relic density, as well as constraints from dark matter (DM) direct detection experiments.\footnote{For related work on the impact of LHC Higgs limits on Higgs portal dark matter, see Refs.~\cite{Hinvisible,Englert:2011yb}, while for more general DM features of Higgs portal models, see Refs.~\cite{Barger:2007nv}.}

For simplicity we consider cases where the DM is either a scalar or a fermion which is a singlet under SM gauge symmetries, and take as free parameters the DM mass, its coupling to the Higgs boson, and  the Higgs mass.
The minimal models describing interactions of the Higgs boson with a scalar and a fermionic DM are \cite{singscalar,Kanemura:2010sh}:
\begin{align}
	{\cal L} &=\delta_c \,m_s^2 |S|^2 + \delta_c\, \lambda_s H^\dagger H |S|^2\,,\\
	{\cal L} &= \delta_c \,m_f \bar\psi \psi + \delta_c\, \frac{\lambda_f}{\Lambda}H^\dagger H \bar\psi \psi \,, 
	 \label{eqn:vector}
\end{align}
where  $\delta_c=1/2$ for a real scalar and a Majorana fermion and 1 otherwise. 

The requirement that the invisible decay width of the Higgs is comparable to the visible decay width sets a lower bound on the couplings:
\be
\lambda_{s} \ , \tilde{\lambda}_f   \; \agt \; {\cal O}\left(1\right) \ ,
\ee
where $\tilde\lambda_f = \lambda_f (v/\Lambda)$ is the effective coupling of the dark fermion to the Higgs scalar. On the other hand, the lack of definitive signals from direct detection experiments of DM places an upper bound on the scattering rate of DM with the nuclei, which depends on the product of two factors: 1) the local density of the DM and 2) the interaction strength of the DM with quarks and gluons inside the nuclei. The local density is inherited from the relic density and is inversely proportional to the thermal average of the DM annihilation rate. Therefore we see the Higgs coupling to DM cancels completely in the signal rate in direct detection experiments, unless additional annihilation channels of the DM exist. We will see that these considerations place strong constraints on the parameter space of minimal models, and satisfying all three conditions: Higgs invisible width, direct detection, and relic density is a non-trivial task. Note however that the relic density constraint could be relaxed, either by allowing the DM particles to annihilate through additional channels, or by letting them decay, either outside of the detector or into final states with large SM backgrounds.

The Higgs decay width is easily obtained in analytic form,
\begin{align}
\Gamma_{ss} & = \delta_c \,{\lambda_s^2 v^2\over 16 \pi m_h}\sqrt{1-{4 m_s^2\over m_h^2}}\,,\\
	\Gamma_{ff} & = \delta_c \,\frac{1}{8 \pi} {\tilde\lambda}_f^2 \, m_h \left(1 - \frac{4 m_f^2}{m_h^2} \right)^{3/2} \,,
\end{align}
The relic density can be obtained from the Higgs mediated annihilation cross section in the nonrelativistic limit \cite{singscalar}:
\begin{align}
	(\sigma v)_{SS\to X_{\rm SM}} & = \frac{2 \lambda_s^2 v^2}{(4 m_s^2 - m_h^2)^2  + m_h^2 \Gamma_h^2} \; \frac{\Gamma_{h\to \rm SM} (m_h = 2 m_s)}{2m_s} \,, \\
	(\sigma v)_ {\psi\psi\to X_{\rm SM}} & = v_{\rm rel}^2\frac{\tilde\lambda_f^2 m_f^2 }{(4 m_f^2 - m_h^2)^2  + m_h^2 \Gamma_h^2} \; \frac{\Gamma_{h\to \rm SM} (m_h = 2 m_f)}{2m_f} \,, 
\end{align}
where $\Gamma_h = \Gamma_{h \to \rm SM} + \Gamma_{h \to ss}$ is the total Higgs width, and $\Gamma_{h\to \rm SM} (m_h = 2 m_s)$ denotes the  width of the Higgs boson decays into SM particles for a Higgs mass of $2m_s$, which is a convenient way of summing over all possible final states via the virtual Higgs exchange for a center-of-mass energy of $2m_s$. The relative velocity of the annihilating particles $v_{\rm rel}$ appears for fermionic DM since the annihilation occurs via $p$-wave Higgs exchange.  Using these formulas, it is easy to incorporate dark matter annihilation to $WW^*$ and to gluon pairs, which are included in our codes. The relic density can now be obtained using the standard approximate solutions to the Boltzmann equations~\cite{Servant:2002aq}:
\begin{align}
	\Omega_{N_1} h^2 \approx \frac{1.04\times 10^9~{\rm GeV}^{-1}}{M_{\rm pl}} \frac{x_F}{\sqrt{g_\star}} \frac{1}{a + 3b/x_F}\,,
\end{align}
where the freeze-out temperature $x_F = M_1/T_F$ is determined numerically from
\begin{align}
	x_F = \log \left[ c(c+2)\sqrt{\frac{45}{8}} \frac{g_d}{2 \pi^3} \frac{M_1 M_{\rm pl} (a + 6b/x_F)}{\sqrt{g_\star}\sqrt{x_F}} \right]\,,
\end{align}
and the annihilation cross section is decomposed as $\langle \sigma v \rangle = a + b v^2$.\footnote{Higher order terms in the velocity expansion would be important near threshold or resonance.} The remaining parameters are the number DM degrees of freedom $g_d$, $g_\star = 92$, the Planck mass $M_{\rm pl} = 1.22\times 10^{19}$~GeV and $c=1/2$. 

 Note that for non-self-conjugate fields such as a complex scalar or a Dirac fermion, $\frac{1}{2} \langle\sigma v\rangle$ must be used to calculate the relic density, which accounts for the fact that a DM particle cannot annihilate with itself, but only with its corresponding anti-particle \cite{Srednicki:1988ce}. 

Finally the event rate at DM direct detection experiments is determined by the elastic DM-nucleon scattering cross section, which is mediated by $t$-channel Higgs exchange, 
\begin{align}
	\sigma_{{\rm el},s} & = \frac{\lambda_s^2 m_N^2 f_N^2}{4 \pi m_h^4} \frac{m_N^2}{(m_s+m_N)^2}\,, \\
	\sigma_{{\rm el},f} & = \frac{\tilde\lambda_f^2 m_N^2 f_N^2}{\pi m_h^4 v^2} \frac{m_N^2 m_f^2}{(m_N+m_f)^2} \,,
\end{align}
where $f_N$ is the effective Higgs-nucleon coupling and $m_N$ is the nucleon mass. The most precise determination of $f_N$ comes from the lattice. Using
\begin{align}
	f_N & = \left(\sum_{u,d,s} f^{N}_q + \frac{6}{27} f^{N}_{G} \right),
\end{align}
and the numerical values given in~\cite{Giedt:2009mr}, we obtain $f_N = 0.32$ for both protons and neutrons.  For scattering rates of more general DM spin, see Ref.~\cite{Barger:2008qd}.

\begin{figure}[t]
\begin{center}
\includegraphics[scale=1.1]{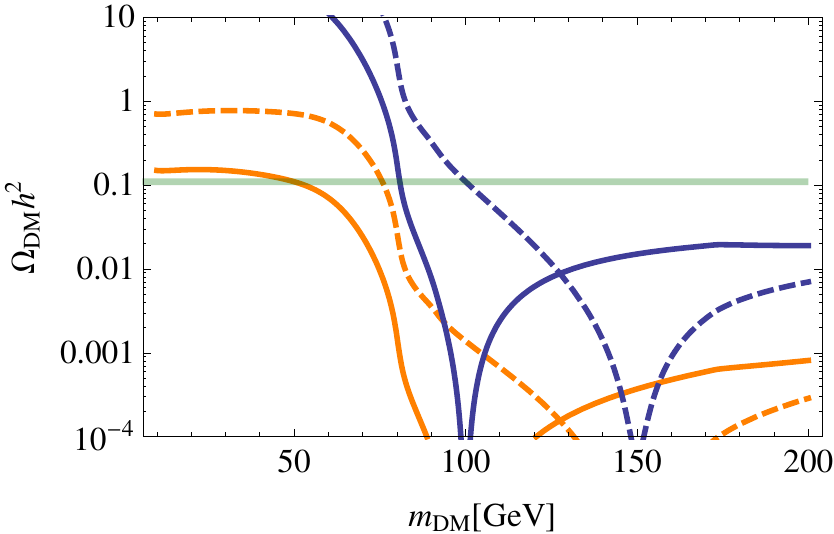} 
\caption{\it Relic density for scalar (orange/light grey) and fermion (blue/dark grey) dark matter as a function of the dark matter mass. The curves are shown for Higgs masses of 200~GeV (solid) and 300~GeV (dashed), for a fixed couplings of $\lambda_s = 1$ and $\tilde\lambda_f = 1$ respectively for scalars and fermions. The light green band is the WMAP-7 measured \cite{Jarosik:2010iu} dark matter relic density. }
\label{fig:rd1}
\end{center}
\end{figure}

In Fig.~\ref{fig:rd1} we show the DM relic density as a function of the mass for benchmark scenarios of $\lambda_s, \tilde{\lambda}_f = 1$. For both a scalar and a fermionic DM the relic density drops significantly around $m_{\rm DM} \sim m_W$, where annihilation into electroweak gauge bosons becomes kinematically allowed. We see that a correct relic density with order unity coupling to the Higgs can be achieved for a scalar DM masses below $m_W$  and a fermionic DM mass above $m_W$. We therefore focus our attention to these two mass ranges.

\begin{figure}[htbp]
\begin{center}
\includegraphics[width=.46\textwidth, angle=0]{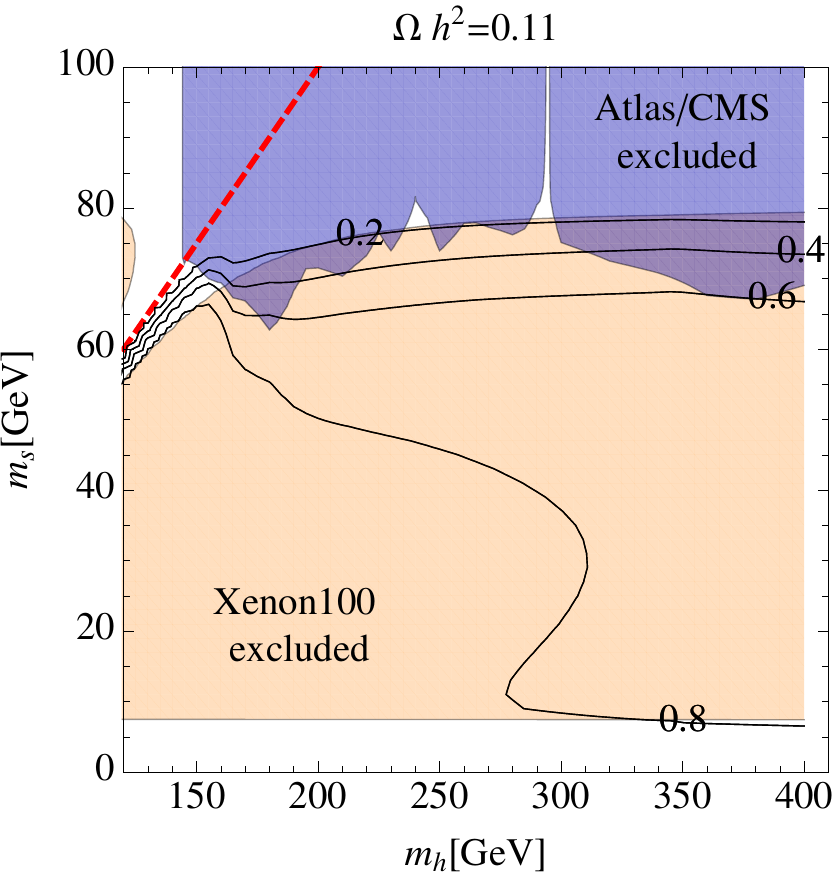}  \hspace*{1cm}
\includegraphics[width=.46\textwidth, angle=0]{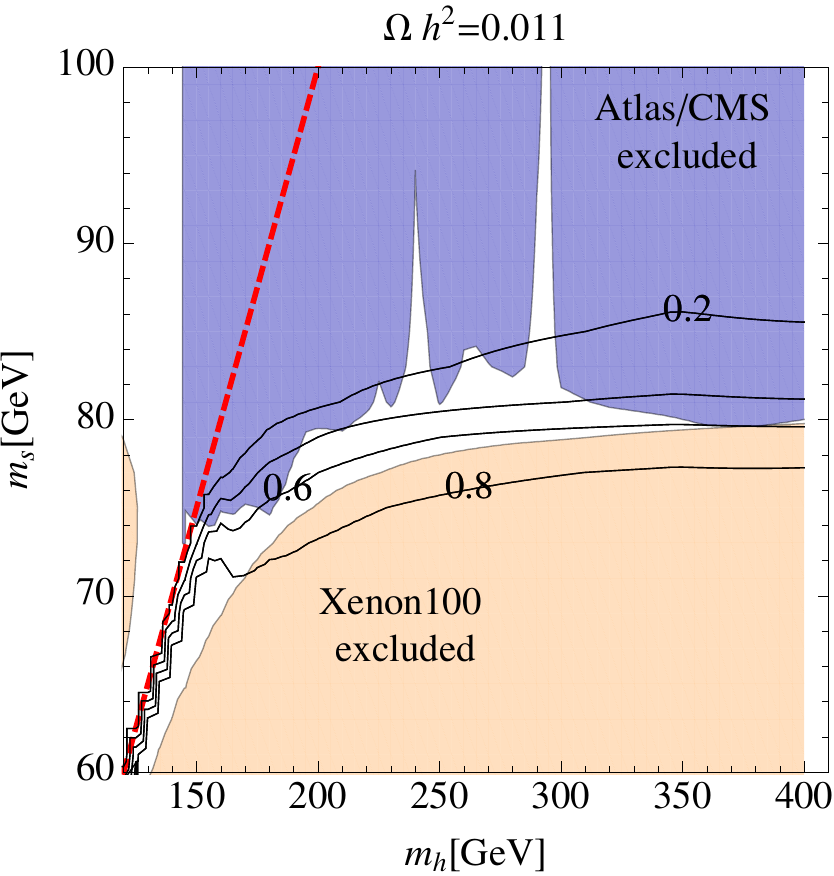}  \\
\includegraphics[width=.46\textwidth, angle=0]{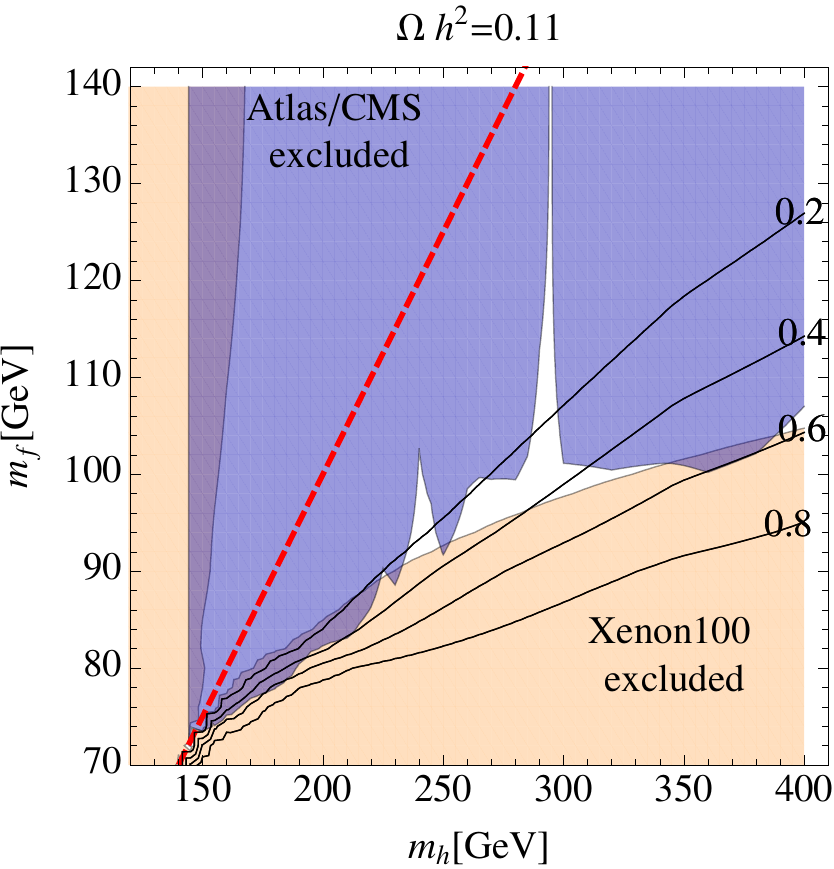}  \hspace*{1cm}
\includegraphics[width=.46\textwidth, angle=0]{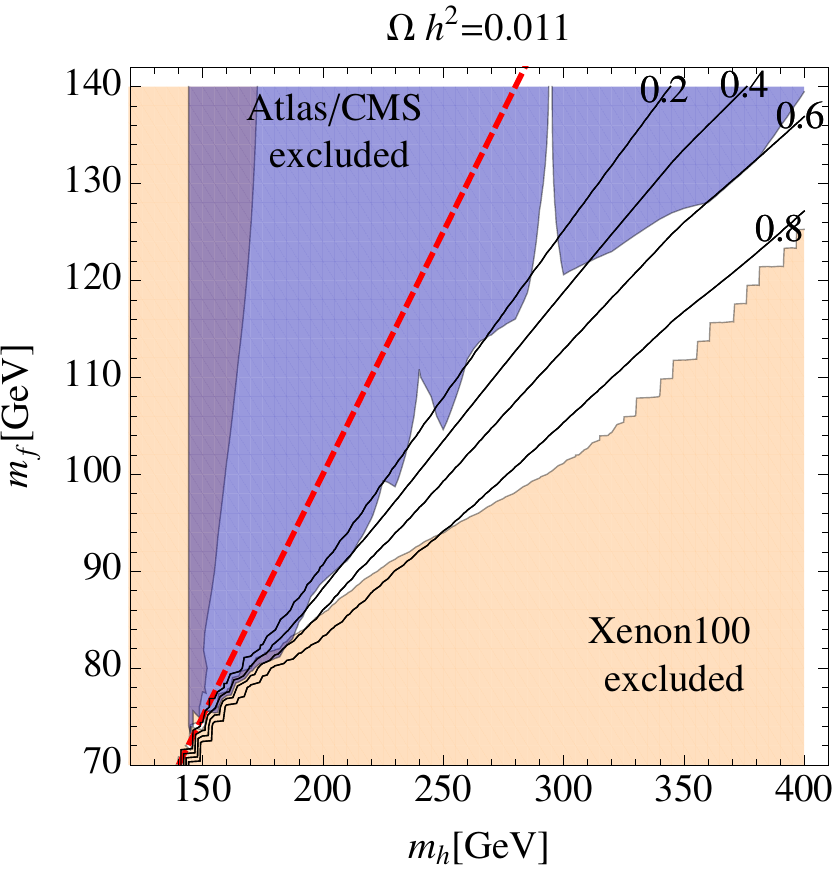}  
\caption{\it Allowed parameter space for minimal scalar DM (top panel) and fermionic DM (bottom panel), in the $(m_h, m_{\rm DM})$ plane. The orange (light grey) shaded region is excluded by direct dark matter searches, while the blue (dark grey) shaded region is excluded by the ATLAS and CMS Higgs search limits. In the left panel the couplings $\lambda_s$ and $\tilde{\lambda}_f$ are fixed by requiring that $\Omega_{\rm DM} h^2 = 0.11$ while for the right panel we require $\Omega_{\rm DM} h^2 = 0.011$. The direct detection rates are rescaled accordingly. The black solid lines represent contours of invisible Higgs branching fraction ranging from 20\% to 80\%.  The red (thick) dotted line gives $m_h=2m_{\rm DM}$, above which the Higgs cannot decay to DM.}
\label{fig:mhms}
\end{center}
\end{figure}

We are now in a position to consider whether there is viable parameter space with $m_h \gtrsim 200$~GeV that could be consistent with the current LHC Higgs limits, the observed relic density, and direct detection constraints on DM. The results are presented in the  $(m_h, m_{\rm DM})$ plane in Fig.~\ref{fig:mhms}, where for each $(m_h, m_{\rm DM})$  we determine the coupling $\lambda$ by the relic density constraint. In particular, we consider cases where the invisible decay product makes up 100\% and 10\% of the observed relic density, respectively. The different mass ranges for a scalar and a fermionic DM are motivated by the relic density considerations in Fig.~\ref{fig:rd1}. Shown in Fig.~\ref{fig:mhms} are : 1) contours of invisible Higgs branching fraction ranging from 20\% to 80\%, 2) limits from Xenon~100  on the spin independent DM nucleon cross section \cite{Aprile:2011hi}, and 3)  ATLAS and CMS limits on $\sigma_{\rm Higgs}/\sigma_{\rm SM}$ re-interpreted as lower bounds on the invisible Higgs branching fraction. The plots shown are for the case of a complex scalar/Dirac fermion, but the limits on the parameter space are very similar for the corresponding cases of real scalar or Majorana fermion DM.

For scalar  DM, the top panel in Fig.~\ref{fig:mhms} suggests that  the minimal scenario where the DM annihilates solely through the virtual Higgs exchange is tightly constrained, except for the region of Higgs masses below $145$~GeV \cite{Hinvisible} or a DM mass below 20 GeV. However, for a light scalar DM in this mass region, there may be large fluxes of anti-protons from the galactic center due to annihilations of DM through the Higgs exchange \cite{Cao:2009uv}, which is severely constrained by the lack of excess in the anti-proton spectrum measured by the PAMELA collaboration \cite{Adriani:2010rc}. On the other hand, above $m_h=200$~GeV there is no region compatible with direct detection limits where the Higgs has a large invisible width. If we relax the relic density constraint, the parameter space opens up,  which is illustrated in the top right plot of Fig.~\ref{fig:mhms}, where the coupling $\lambda_s$ is chosen such that the relic density is 10\% of the total observed dark matter density in the universe. Note that for this case we have changed the mass range for $m_s$ to avoid regions where the coupling needs to be nonperturbative. 

For fermions, a similar picture emerges, as is shown in the bottom panel in Fig.~\ref{fig:mhms}. However, compared to the case of a scalar DM, there is slightly more viable region of parameter space where the invisible Higgs width is sizable and the DM annihilates completely through the Higgs exchange. This region will be probed in the near future by the LHC experiments. Relaxing the relic density constraint, a large region of parameter space opens up where the invisible branching fraction of the Higgs is larger than the SM width of the Higgs.

\section{Conclusions}\label{sect:conclude}

Searches for a SM Higgs boson at the LHC have put stringent limits on the allowed range of Higgs masses, essentially excluding a SM-like Higgs boson in the mass range of $145$~GeV$<m_h<450$~GeV. Higgs masses in this range are only viable if the signal rates in the relevant search channels are suppressed. In this work we considered the possibility that such a suppression is due to the dark side of the Higgs boson, where the invisible decay width of the Higgs is comparable to the visible decay width.

We also proposed a method to infer the invisible decay of the Higgs by measuring the total width of the Higgs boson in its decay to four charged leptons. This measurement is possible for Higgs masses above $190$ GeV, where the width of the Higgs boson is comparable to the experimental resolution. Compared to previous studies on width measurements, we have combined the $4\mu$ and the $e^+e^-\mu^+\mu^-$ channels and taken into account the reduced sensitivity due to the decreased signal rate in the presence of invisible Higgs decay modes.  We find that the Higgs width measurement can probe invisible Higgs decays with a better sensitivity than the VBF channel for Higgs masses above 200~GeV. Assuming a large invisible Higgs width, $\Gamma_{\rm inv} \agt \Gamma_{\rm SM}$, a SM total width can be rejected at the 14~TeV LHC with less than 40~fb$^{-1}$ of integrated luminosity for most Higgs masses, while a similar measurement at the current 7~TeV run of the LHC would require more than 100~fb$^{-1}$ of luminosity. 

The combination of total width and event rate measurements could help determine whether the reduction in the signal strength is due a smaller production cross section or a larger total width.

The simplest extension of the SM that results in a large invisible Higgs width is the addition of a gauge singlet DM particle that couples to the SM only through the Higgs boson. In this case, a  correct relic density and a large invisible decay width can be obtained for an order unity coupling. However, when confronting the simplest models with LHC Higgs limits and direct detection constraints, we find that both a scalar and a fermionic DM singlet is heavily constrained. Relaxing the relic density constraint, e.g. by assuming that the DM singlet is only one component of the total dark matter in the universe, we find sizable regions of viable parameter space. These scenarios will be probed in the near future by the ongoing Higgs and DM searches.

\begin {acknowledgements}
P.~S. would like to thank T.~Gehrmann for useful discussions. I.~L., P.~S., and C.~W. thank lively atmosphere at the CERN TH-LPCC summer institute on LHC physics, where part of this work was performed. This work is supported in part by U.S. Department of Energy under grant numbers DE-AC02-06CH11357, DE-FG02-84ER40173, DE-FG02-91ER40684 and DE-FG02-95ER40896. \end{acknowledgements}

\end{document}